# RICH METHANE PREMIXED LAMINAR FLAMES DOPED BY LIGHT UNSATURATED HYDROCARBONS

# PART III: CYCLOPENTENE

H.A.GUENICHE, P.A. GLAUDE*, R. FOURNET, F. BATTIN-LECLERC

Département de Chimie-Physique des Réactions,
UMR 7630 CNRS-INPL, Nancy-University,
1 rue Grandville, BP 20451, 54001 NANCY Cedex, France

SHORTENED RUNNING TITLE : **METHANE FLAMES DOPED BY CYCLOPENTENE**

* E-mail : pierre-alexandre.glaude@ensic.inpl-nancy.fr ; Tel.: 33 3 83 17 51 01 , Fax : 33 3 83 37 81 20

In line with the studies presented in the parts I and II of this paper, the structure of a laminar rich premixed methane flame doped with cyclopentene has been investigated. The gases of this flame contains 15.3% (molar) of methane, 26.7% of oxygen and 2.4% cyclopentene corresponding to an equivalence ratio of 1.79 and a ratio $C_5H_8$ / $CH_4$ of 16 %. The flame has been stabilized on a burner at a pressure of 6.7 kPa using argon as dilutant, with a gas velocity at the burner of 36 cm/s at 333 K. The temperature ranged from 627 K close to the burner up to 2027 K. Quantified species included usual methane $C_0$-$C_2$ combustion products, but also propyne, allene, propene, propane, 1-butene, 1,3-butadiene, 1,2-butadiene, vinylacetylene, diacetylene, cyclopentadiene, 1,3-pentadiene, benzene and toluene. A new mechanism for the oxidation of cyclopentene has been proposed. The main reaction pathways of consumption of cyclopentene and of formation of benzene and toluene have been derived from flow rate analyses.

## INTRODUCTION

While the formation of PAHs and soot represents an important area of interest for kineticists for two decades, questions still remain even concerning the formation and the oxidation of the first aromatic compounds during the combustion of large hydrocarbons, especially [1]. In the literature, the formation of benzene is mostly related to the reactions of $C_2$ (acetylene), $C_3$ or $C_4$ unsaturated species [2-6]. Nevertheless, some papers suggest a link between $C_6$ and $C_5$ cyclic species [1,7,8]. In a first part of this paper [9], we have investigated the reactions of allene and propyne, as they are precursors of propargyl radicals, which have an important role in forming benzene. In a second part [10], we have analyzed the reaction of 1,3-butadiene in a methane flame and putted in evidence an important production of benzene due to reactions of $C_4$ compounds under these particular conditions. To finish this work, it is interesting to study the reactions of cyclopentene, which is a source of $C_5$ radicals.

The oxidation of cyclopentene has already been experimentally investigated in a shock tube [11] and in laminar premixed flames with cyclopentene as only fuel [12, 13]. Lindstedt and Rizos [14] have proposed a model to simulate the results obtained in flames by Lamprecht et al. [12].

Using the same methodology and similar experimental conditions as in parts I and II, the purpose of the present study is to experimentally investigate the structure of a premixed laminar methane flame containing cyclopentene. That will allow comparisons with the structures of the pure methane flame and the flames doped by allene, propyne and 1,3-butadiene. These results have been used to develop a new mechanism for the oxidation of cyclopentene.

## EXPERIMENTAL RESULTS

In line with our previous studies [9, 10], a laminar premixed flat flame has been stabilized on the burner at 6.7 kPa (50 Torr) with a gas flow rate of 3.32 l/min corresponding to a gas velocity at the burner of 36 cm/s at 333 K and mixtures containing 55.6% argon, 15.3% methane (99.95 %, pure supplied by Alphagaz - L'Air Liquide), 26.7% oxygen and 2.4% cyclopentene corresponding to an equivalence ratio of 1.79. Methane (99.95 % pure) was supplied by Alphagaz - L'Air Liquide. Oxygen (99.5% pure) and argon (99.995% pure) were obtained from Messer and liquid cyclopentene was purchased from Fluka, with a purity of 99%.

Problems of stability of the flame explain why a slightly higher dilution (55.6% argon instead of 42.9%) has been used compared to our previous flames containing all 20.9% of methane [9, 10]. As we have used the same apparatus, the same method to measure temperature and the same analytical techniques as what is extensively described in the part I of this paper [9], they are not presented again here. A single improvement was made to allow a liquid reactant to be injected in the gaseous mixture feeding the flame. Liquid cyclopentene was contained in a glass vessel pressurized with argon. After each load of the vessel, argon bubbling and vacuum pumping were performed in order to remove oxygen traces dissolved in the liquid hydrocarbon fuel. The liquid reactant flow rate was controlled by using a liquid mass flow controller, mixed to the carrier gas and then evaporated by passing through a single pass heat exchanger whose temperature was set above the boiling point of the mixture. Carrier gas flow rate was controlled by a gas mass flow controller located before the mixing chamber.

Aromatic (benzene and toluene) species and linear species containing 4 or less atoms or carbons were common to those observed in the flames doped by allene, propyne [9] and 1,3-butadiene [10], apart from the absence of butynes. $C_5$ compounds were also analysed using the gas chromatograph fitted with a Haysep packed column, a flame ionisation dectector and nitrogen as gas carrier gas. The identification of these compounds was performed using GC/MS and by comparison of retention times when injecting the product alone in gas phase. Figure 1 presents a typical chromatogram of $C_1$-$C_6$ compounds obtained for the flame doped with cyclopentene. The observed $C_5$ compounds were cyclopentene, which was the reactant, cyclopentadiene and 1,3-pentadiene. Unfortunately, The peaks of cyclopentene and cyclopentadiene cannot be distinguished. Compounds containing more than 7 atoms of carbon, such as naphthalene, could not be analysed with our apparatus.

FIGURE 1

Water and small oxygenated compounds, such as acetaldehyde, formaldehyde, acroleine, were detected by GC-MS but not quantitatively analysed.

Figure 2 displays the experimental temperature profiles obtained with and without the probe showing as previously that the presence of the probe induces a thermal perturbation involving a lower temperature. Without the probe, the lowest temperatures measured the closest to the burner were around 600 K. The highest temperatures were reached between 0.68 and 0.89 cm above the burner and were around 2020 K.

FIGURE 2

Figures 3 and 4 present the profiles of the $C_0$-$C_2$ species involved in the combustion of methane vs. the height above the burner. The profile of carbon dioxide (fig. 3d) shows a marked inflexion point as in our previous doped flames.

FIGURES 3 AND 4

Contrary to the four previous flames [9, 10], ethylene (fig. 4c) is produced first and reaches its maximum concentration close to the burner, around 0.5 cm. The profiles of ethane (fig. 4d) and of acetylene (fig. 4b) peak around 0.6 cm, and that of carbon monoxide around 1.3 cm. While the maximum value reached by the mole fraction of ethane is slightly decreased by the addition of cyclopentene, those of ethylene (0.006 compared to 0.002 in the pure methane flame) and acetylene (0.012 compared to 0.001) are strongly increased, but are close to the value obtained with the addition of 1,3-butadiene [10].

Figure 5 presents the profiles of the observed $C_3$ products, which all peak around 0.5 cm above the burner. While the experimental formation of propane (fig. 5d) is close to that simulated in the pure methane flame, the formation of allene (fig. 5a), propyne (fig. 5b) and propene (fig. 5c) are much increased by the presence of the additive. The amounts of $C_3$ species is close to what was observed in the flame doped by 1,3-butadiene [10].

FIGURE 5

Figure 6 present the profiles of $C_4$ species. 1-butene (fig. 6a) and 1-3 butadiene (fig. 6b) are produced first and reaches their maximum concentration close to the burner, around 0.5 cm. The profiles of 1,2-butadiene (fig. 6c) and vinylacetylene (fig. 6d) peak around 0.55 cm and that of diacetylene (fig. 6e) around 0.6 cm. While the amount of diacetylene is almost unchanged, the maximum of the peak of 1,2-butadiene is much lower and that of vinylacetylene much higher than what was observed in the flame doped by 1,3-butadiene [10].

FIGURE 6

Figure 7 presents the profiles of $C_5$ species. As methane, cyclopentene (Fig. 7a) is consumed in the first stage of the flame, but the total consumption of the $C_5$ cyclic species occurs closer to the burner, at 0.5 cm height, while some methane (Fig. 3a) remains up to 0.7 cm. The profile of 1,3-pentadiene peaks around 0.5 cm from the burner. Figure 7 presents also the profiles of benzene (fig. 7c) and toluene (fig. 7d). The maximum of the peak of benzene is obtained around 0.4 cm above the burner and that of toluene around 0.6 cm. The maximum value reached by the mole fraction of benzene is around three times that measured in the flame doped by 1,3-butadiene [10] and ten times that measured in the flame doped with allene [9]. The amount of toluene is more than fifty times higher than in the flame doped by 1,3-butadiene.

FIGURE 7

## DESCRIPTION OF THE PROPOSED MECHANISM

The mechanism proposed here to model the oxidation of cyclopentene includes the previous mechanisms that were built to model the oxidation of $C_3$-$C_4$ unsaturated hydrocarbons [9, 10, 15, 16], benzene [17] and toluene [18]. Thermochemical data were estimated by the software THERGAS developed in our laboratory [19], which is based on the additivity methods proposed by Benson [20].

*Reaction base for the oxidation of $C_3$-$C_4$ unsaturated hydrocarbons [9, 10]*

This $C_3$-$C_4$ reaction base, which is described in details in the previous parts of this paper, was built from a review of the literature and is an extension of our previous $C_0$-$C_2$ reaction base [21]. This $C_0$-$C_2$ reaction base includes all the unimolecular or bimolecular reactions involving radicals or molecules including carbon, hydrogen and oxygen atoms and containing less than three carbon atoms. The kinetic data used in this base were taken from the literature and are mainly those proposed by Baulch *et al.* [22] and Tsang *et al.* [23]. The $C_0$-$C_2$ reaction base was first presented in the paper of Barbé *et al.* [21] and has been later up-dated [15].

The $C_3$-$C_4$ reaction base includes reactions involving $C_3H_2$ , $C_3H_3$ , $C_3H_4$ (allene and propyne), $C_3H_5$,$C_3H_6$, $C_4H_2$, $C_4H_3$ , $C_4H_4$, $C_4H_5$, $C_4H_6$ (1,3-butadiene, 1,2-butadiene, methyl-cyclopropene, 1-butyne and 2-butyne), $C_4H_7$ (6 isomers), as well as some reactions for linear and branched $C_5$ compounds and the formation of benzene. It has been comprehensively described in the two first parts of this paper [9, 10].

In this reaction base, pressure-dependent rate constants follow the formalism proposed by Troe [24] and efficiency coefficients have been included. It has been validated by modeling experimental results obtained in a jet-stirred reactor for methane and ethane [21], profiles in laminar flames of methane, acetylene and 1,3-butadiene [15] and shock tube autoignition delay times for acetylene, propyne, allene, 1,3-butadiene [15], 1-butyne and 2-butyne [16]. An improved version has recently been used to model the structure of a laminar premixed flame of methane doped with allene, propyne and 1,3-butadiene [9, 10].

*Mechanisms for the oxidation of benzene and toluene*

Our mechanism for the oxidation of benzene contains 135 reactions and includes the reactions of benzene and of cyclohexadienyl, phenyl, phenylperoxy, phenoxy, hydroxyphenoxy, cyclopentadienyl, cyclopentadienoxy and hydroxycyclopentadienyl free radicals, as well as the reactions of ortho-benzoquinone, phenol, cyclopentadiene, cyclopentadienone and vinylketene, which are the primary products yielded [17]. Validations have been made using for comparison experimental results obtained in a jet-stirred [17] and a plug flow [25] reactors, profiles in a laminar flame of benzene [26] and shock tube autoignition delay times [17].

The mechanism for the oxidation of toluene contains 193 reactions and includes the reactions of toluene and of benzyl, tolyl, peroxybenzyl (methylphenyl), alcoxybenzyl and cresoxy free radicals, as well as the reactions of benzaldehyde, benzyl hydroperoxyde, cresol, benzylalcohol, ethylbenzene, styrene and bibenzyl [18]. Validations have been made using for comparison experimental results obtained in a jet-stirred [18] and a plug flow [25] reactors, and shock tube autoignition delay times [18].

*Mechanism proposed for the oxidation of  cyclopentene*

The part of mechanism, described below and given in Table I, is included in a file which also contains mechanisms described above and which can be used to run simulations using CHEMKIN [27]. Table II presents the names, the formulae and the heats of formation of the species specific to this mechanism.

TABLES I AND II

We have considered the unimolecular reactions of cyclopentene, the additions to the double bonds and the H-abstractions by oxygen molecules, hydrogen and oxygen atoms and hydroxyl, methyl and ethyl radicals. Unimolecular reactions include molecular dehydrogenation (reaction 1 in Table I) with a rate constant proposed by Lewis *et al.* [28], decomposition by transfer of a H atom and isomerization to yield 1,2-pentadiene (reaction 2), which leads to 1,3-pentadiene (reaction 3) and to propyne and ethylene (reaction 4). The reactions of 1,2-pentadiene have been studied by Herzler *et al.* [29]. The rate constants of the unimolecular decompositions to give H atoms and cyclopentenyl radicals (reactions 5-7) have been deduced from that of the reverse reaction, k = 1.0 $x10^{14}$ $s^{-1}$ according to Allara *et al.* [31]. We have written the addition to the double bond of H-atoms to give cyclopentyl radicals (reaction 11) and the addition of OH radicals (reaction 12) to give ethylene and HCO radicals. The bimolecular initiations (reactions 8-10) and the H-abstractions (reactions 13-26) involved the formation of the three possible types of cyclopentenyl radicals (alkylic C5H7#, allylic C5H7#Y and vinylic C5H7#V). The rate constants of the bimolecular initiations with oxygen molecule have been calculated as proposed by Ingham *et al.* [32]. The rate constants for the abstractions of alkylic H-atoms were deduced from the correlations proposed by Buda *et al.* for alkanes [34] and the rate parameters for the additions on the double bond and the abstractions of allylic and vinylic H-atoms were obtained from the structure-reactivity relationships proposed by Heyberger *et al.* for alkenes [33, 35]. For the addition of H-atoms, the high pressure limit value calculated by Sirjean *et al.* [30] by quantum calculations for the reverse reaction leads to a overestimation of the formation of ethylene. That can be explained by a fall-off effect at the low pressure used for the present study.

The reactions of cyclopentenyl radicals involved isomerizations (reactions 27, 28),

decompositions by breaking of a C-C bond to form linear $C_5$ radicals including two double bonds (reactions 29, 31, 40) or a triple bond (reaction 39), the formation of cyclopentadiene by decompositions by breaking of a C-H bond (reactions 43, 44) or by oxidation with oxygen molecules (reactions 45, 46) and terminations steps (reactions 47, 50). Termination steps were written only for the resonance stabilized cyclopentenyl radicals: disproprtionnations with H-atoms and OH radicals gave cyclopentadiene, combinations with $HO_2$ radicals led to ethylene and $CH_2CHCO$ and OH radicals, and combinations with $CH_3$ radicals formed methylcyclopentene. The isomerizations (for the radical stabilized isomer, reactions 32-35) and the decompositions by breaking of a C-C bond of the linear $C_5$ radicals including two double bonds (reactions 30, 36-38, 42), or a triple bond (reaction 41) were also written, while those by breaking of a C-H bond were not considered. The rate parameters of the isomerizations and the decompositions of cyclic and linear radicals were estimated using the structure-reactivity relationships proposed for alkenes by Heyberger et al. [33, 35]. The activation energies of the isomerizations of cyclic radicals have been increased by 5 kcal/mol to take into account the formation of a bicyclic transition state. In the case of the decompositions, the A-factors and activation energies depended on the types of reacting radicals, bonds broken and products obtained. As previous work [30] have shown that the ring strain energy is not recovered during the opening of a cycle by beta-scission decomposition, the rate constants of these reactions have been taken equal to that of the similar beta-scission decompositions of linear radicals.

The decomposition by breaking of a C-H bond of cyclopentyl radicals (reaction 51) led to the formation of 1-penten-5-yl radicals with a rate constant proposed by Sirjean et al. [30]. These linear radical can isomerize to give the resonance stabilized 1-penten-3-yl radicals (reaction 52), the reactions of which were considered in the $C_3$-$C_4$ reaction base [9], decompose to yield ethylene and allyl radicals (reactions 53), or pentadiene and a H atom (reactions 54) or react by oxidation with oxygen molecules (reactions 55).

The reactions of cyclopentadiene were part of the mechanism for the oxidation of benzene, but reactions for the consumption of methylcyclopentene and methylcyclopentadiene, obtained by recombination of cyclopentadienyl and methyl radicals (reaction 73), had to be added. The H-abstractions by H-atoms (reactions 56-59 and 76-77) and OH radicals (reactions 60-63 and 78-79) have been written, but the abstractions of vinylic H-atoms have been neglected. The addition of H-atoms to methylcyclopentadiene have also been considered (reactions 74-75). The reactions of the radicals deriving from methylcyclopentene included decompositions by breaking of a C-C bond (reactions 64-66), oxidation with oxygen molecules to give methylcyclopentadiene and $HO_2$ radicals (reactions 69-70) and combinations of the two allylic radicals with H-atoms (reactions 71-72). For the radicals deriving from methylcyclopentadiene, we have considered, as proposed by Lifshitz et al. [8], the isomerization between both radicals (reaction 80) and the formation of cyclohexadienyl radicals, which are precursors of benzene, from cyclopentadienyl methylene radicals (reaction 81). We have not considered the formation of cyclohexadienyl radicals from the resonance stabilized methyl cyclopentadienyl radicals, as proposed by Marinov et al. [5], but their recombination with H-atoms (reaction 82) was also added.

In order to reproduce the important formation of toluene experimentally observed, we have considered the formation of benzyl radicals from the addition of acetylene to cyclopendadienyl radicals (reaction 83) with a rate constant about three times higher than what is usually considered for such an addition [33, 35]. While the decomposition of benzyl radicals to form acetylene and cyclopendadienyl radicals, which was considered in our mechanism for the oxidation of toluene, has been studied by several authors [39-41] and has been shown to occur through a several steps mechanism, the reverse addition has never been directly investigated.

## COMPARISON BETWEEN EXPERIMENTAL AND SIMULATED RESULTS

Simulations were performed using PREMIX from CHEMKIN II [27]. To compensate the perturbations induced by the quartz probe and the thermocouple, the temperature profile used in calculations is an average between the experimental profiles measured with and without the quartz probe, shifted 0.15 cm away from the burner surface, as shown in figure 2.

Figures 3 and 4 show that the model reproduces satisfactorily the consumption of reactants and the formation of the main $C_0$-$C_2$ products related to the consumption of methane in the flame doped with cyclopentene. Only the formation of ethane is underpredicted by a factor almost 2.

To decouple the effect due to the increase of equivalence ratio ($\phi$) and that induced by the presence of cyclopentene, figures 3 and 4 display also the results of a simulation performed for a flame containing 15.3% methane and 23.0% oxygen (with no $C_5$ additive) for $\phi$= 1.8, i.e. equal to that of the doped flame. As the temperature rise is mainly influenced by $\phi$, we have used the same temperature profile as to model the doped flame. As in our previous studies, the profiles of methane are very similar for the doped and undoped flames at the same $\phi$. The profiles of oxygen, carbon oxides and hydrogen are different in the doped and pure methane flames at $\Phi$=1.8 due to a difference in the C/O and C/H ratios. At the same equivalence ratio, the maximum of ethane mole fraction is lower by almost a factor 2 in the doped flame than in the pure methane flame; that could indicates that the underestimation obtained by simulation could be due to a deficit in methyl radicals, the recombination of which is the major channel of formation of ethane. The maximum of ethylene mole fraction is slighted increased by the addition of cyclopentene, while that of acetylene is multiplied by a factor 2.5.

Figures 5 and 6 present the comparison between experimental and simulated data for $C_3$ and $C_4$ species respectively. The formation of unsaturated $C_3$ compounds are modeled within a factor around 2, but the formation of propane is underestimated by a factor up to 3. Comparison with the simulation of a pure methane flame at Φ=1.8 shows that allene, propyne and propene are much more abundant (factors around 10) in the doped flame, while the maximum of propane mole fraction is lower by a factor around 4. Specific reactions leading to unsaturated $C_3$ products are then also induced by the presence of the cyclic additive. The decrease of the concentration of propane in the doped flame could confirm a possible deficit in methyl radicals in our simulations. The formation of 1-butene, 1,3-butadiene, vinylacetylene and diacetylene are satisfactorily modeled, while that of 1,2-butadiene is underestimated by a factor 4.

Figure 7 displays the profiles of $C_5$ and aromatic compounds. The consumption of the cyclic $C_5$ compounds is well reproduced, as well as the formation of 1,3-pentadiene and benzene. Toluene is underestimated by a factor about three.

## DISCUSSION

Figure 8 displays the flows of consumption of the cyclopentene at a temperature about 990 K corresponding to a conversion of 47%. Under these conditions, cyclopentene is mainly consumed by additions of H-atoms (reaction 11 in Table I, 53 % if its consumption) to give cyclopentyl radicals and of OH radicals (reaction 12, 3%) to form ethylene and HCO radicals and by H-abstraction by H atoms and OH radicals to give the resonance stabilized cyclopentenyl radicals (C5H7#Y in table II)  (reactions 14 and 17, 35%), the alkenyl cyclopentenyl radicals (C5H7#)  (reactions 13 and 16, 5%) and the vinylic cyclopentenyl radicals (C5H7#V)  (reactions 15 and 18, 1.5%).

FIGURE 8

Cyclopentyl radicals isomerize rapidly to form 1-penten-5yl radicals. These linear radicals are completely decomposed to give ethylene and resonance stabilized allyl radicals (reaction 53), which react mainly by termination steps. Their combination and their disproportionation with H-atoms are the main ways of formation of propene and allene, respectively. The combination of allyl radicals with methyl radicals leads to 1-butene and that with $HO_2$ to acrolein, the reactions of which involve the formation of ethylene and HCO radicals or carbon monoxide and vinyl radicals. The rate constant of the additions of H-atoms to cyclopentene is a sensitive parameter for the formation of ethylene, propene and allene. The fact that the addition of H-atoms to cyclopentene traps these reactive radicals to form resonance stabilized allyl radicals makes the concentration of radicals to be lower at the beginning of the reactive zone in the doped flame than in the pure methane flame as shown in fig. 9a. That explains why the concentration of ethane and propane are much lower in the doped flame.

FIGURE 9

Resonance stabilized cyclopentenyl radicals react mainly by combination with hydrogen atoms to give back cyclopentene (reaction 6), with methyl radicals to produce methyl cyclopentene (reaction 50) and with $HO_2$ radicals to form ethylene and $CH_2CHCO$ and OH radicals (reaction 47). Minor channels include the breaking of a C-H bond (reaction 44) and the oxidation with oxygen molecules (reaction 46) to give cyclopentadiene and the opening of the cycle (reaction 31), which is the main source of 1,3-pentadiene. The almost only reaction of alkenyl cyclopentenyl radicals is their decomposition to give cyclopentadiene and H-atoms (reaction 43). Vinylic cyclopentenyl radicals react by the opening of the cycle (reaction 39) leading ultimately to ethylene and propargyl radicals (reaction 40) which are a source of propyne and aromatic compounds.

As shown in figure 9b, cyclopentadiene is an important product of this doped flame and reacts mainly by H-abstraction by H atoms and OH radicals to give the resonance stabilized cyclopentadienyl radicals, which react mainly by termination steps. Their combination with methyl radicals produce methylcyclopentadiene (reaction 73), which is the major source of benzene. Their self-combination leads to bicyclopentadienyl, which could rearrange to give naphthalene, which was not considered here. Their combination with $HO_2$ radicals give OH and cyclopentadionyl radicals, which decompose to give HCO and acetylene or H-atoms and cyclopentadienone, and that with OH radicals lead to carbon monoxide and 1,3-butadiene. The combinations of cyclopentadienyl radicals with oxygenated radicals are the main sources of acetylene and 1,3-butadiene. The H-abstraction by H-atoms from cyclopentadione leads to vinylacetylene. Another important reaction of cyclopentadienyl radicals is their addition to acetylene (reaction 83), which is a source of toluene. As displayed in Fig. 9b, simulation show the formation of some important cyclic species, which were not detected in our analyses, i.e. methyl cyclopentene, methylcyclopentadiene and cyclopentadienone.

Figure 10 presents a flow rate analysis for the production and consumption of aromatic compounds at a distance of 0.48 cm from the burner corresponding to a temperature of 1350 K. The major source of benzene is the isomerization of cyclopentadienyl methylene radicals (reaction 81), which are obtained either directly by H-abstractions by H-atoms or OH radicals from methylcyclopentadiene (reactions 76 and 77) or by isomerisation from the resonance stabilized methyl cyclopentadienyl radicals (reaction 80), which are also obtained by H-abstractions by H-atoms or OH radicals from methylcyclopentadiene (reactions 78 and 79). Figure 7c shows that simulations using a mechanism in which reactions 81 has been removed lead to a maximum of benzene 3 times lower than with the full mechanism. Benzene can also be obtained by ipso-addition of H-atoms to toluene or ethylbenzene, involving the elimination of methyl or ethyl radicals, respectively. Benzene is mainly consumed by H-abstractions by H-atoms or OH radicals to give phenyl radicals or by addition of O-atoms to produce phenoxy radicals, which lead to phenol or decompose to form carbon monoxide and cyclopentadienyl radicals. Other sources of phenyl radicals include the combination of propargyl radicals, reactions of benzyl radicals with O-atoms, involving the elimination of HCO radicals and the decomposition of benzaldehyde. The major reaction of phenyl radicals is with oxygen and lead to benzoquinone or to phenoxy radicals. The combination of phenyl radicals with methyl radicals is a very minor source of toluene. The second important source of aromatic compounds is the addition of cyclopentadienyl radicals to acetylene (reaction 83), which leads to benzyl radicals. These resonance stabilized radicals react mainly by combination with H-atoms to form toluene, with methyl radical to produce ethylbenzene or with O-atoms to give benzaldehyde. Simulations using a mechanism in which only reaction 83 has been removed (see Fig. 7c and 7d) show that the addition of cyclopentadienyl radicals to acetylene is the most important way to give toluene and is also of importance for the formation benzene, as one way of consumption of toluene is the ipso-addition of H-atoms to give benzene and methyl radicals. The major channel of consumption of toluene it by H-abstraction by H-atoms and OH radicals to give benzyl radicals.

FIGURE 10

## CONCLUSION

This paper presents new experimental results for a rich premixed laminar flame of methane seeded with cyclopentene, as well as a new mechanism developed to reproduce the combustion of this unsaturated cyclic species. Profiles of temperature have been measured and profiles of stable species have been obtained for 21 products, including benzene, toluene and $C_3$, $C_4$ and $C_5$ compounds. The presence of cyclopentene promotes the formation of ethylene and acetylene. The increase of the formation of this first $C_2$ compound is due to the decomposition of cyclopentenyl radical obtained by addition of H-atoms to cyclopentene and the increase in the formation of the second one can be attributed to combination of cyclopentadienyl and $HO_2$ radicals. Cyclopentadiene is the main cyclic species obtained and the formation of the resonance stabilized cyclopentadienyl radicals is then very important.

This important concentration of cyclopentadienyl radicals is responsible for the formation of benzene and toluene, which cannot be detected in the pure methane flame. The use of methane as the background and consequently of a flame rich in methyl radicals favors the formation of methylcyclopentadiene and then the production of benzene from the derived cyclopentadienyl methylene radicals. This flame is then another example of a flame in which the formation of benzene through the $C_3$ pathway is not preponderant, in contrary to most cases of the literature [42]. This study has also allowed us to suggest a potential formation of toluene through the addition of cyclopentadienyl radicals to acetylene.

**TABLE I: REACTIONS OF CYCLOPENTENE AND OF DERIVED $C_5$ AND $C_6$ SPECIES**

The rate constants are given ($k=A\ T^n \exp(-E_a/RT)$) in cc, mol, s, kcal units. Reference numbers are given in brackets when they appear for the first time.

| Reactions | A | n | $E_a$ | References | No |
|---|---|---|---|---|---|
| **Reactions of cyclopentene** | | | | | |
| C5H8#=C5H6#+H2 | $2.2x10^{13}$ | 0.0 | 60.0 | LEWIS74[28] | (1) |
| C5H8-12=C5H8# | $1.8x10^{11}$ | 0.0 | 51.9 | HERZLER01[29] | (2) |
| C5H8-12=C5H8 | $2.2x10^{14}$ | 0.0 | 67.4 | HERZLER01 | (3) |
| C5H8-12=C2H4+pC3H4 | $6.6x10^{12}$ | 0.0 | 58.1 | HERZLER01 | (4) |
| C5H7#+H=C5H8# | $1.0x10^{14}$ | 0.0 | 0.0 | Estimated [a] | (5) |
| C5H7#Y+H=C5H8# | $1.0x10^{14}$ | 0.0 | 0.0 | Estimated [a] | (6) |
| C5H7#V+H=C5H8# | $1.0x10^{14}$ | 0.0 | 0.0 | Estimated [a] | (7) |
| C5H8#+O2=C5H7#+HO2 | $1.4x10^{13}$ | 0.0 | 47.6 | Estimated [b] | (8) |
| C5H8#+O2=C5H7#Y+HO2 | $2.8x10^{12}$ | 0.0 | 35.6 | Estimated [b] | (9) |
| C5H8#+O2=C5H7#V+HO2 | $4.0\ x10^{12}$ | 0.0 | 55.6 | Estimated [b] | (10) |
| C5H8#+H= C5H9# | $1.0x10^{14}$ | 0.0 | 3.2 | Estimated [c] | (11) |
| C5H8#+OH=2C2H4+CHO | $2.7x10^{12}$ | 0.0 | -1.1 | Estimated [d] | (12) |
| C5H8#+H=C5H7#+H2 | $9.0x10^{6}$ | 2.0 | 5.0 | Estimated [e] | (13) |
| C5H8#+H=C5H7#Y+H2 | $1.08x10^{5}$ | 2.5 | -1.9 | Estimated [d] | (14) |
| C5H8#+H=C5H7#V+H2 | $8.2x10^{5}$ | 2.5 | 9.8 | Estimated [d] | (15) |
| C5H8#+OH=C5H7#+H2O | $2.6x10^{6}$ | 2.0 | -0.8 | Estimated [e] | (16) |
| C5H8#+OH=C5H7#Y+H2O | $6.0x10^{6}$ | 2.0 | -1.5 | Estimated [d] | (17) |
| C5H8#+OH=C5H7#V+H2O | $2.2x10^{6}$ | 2.0 | 1.4 | Estimated [d] | (18) |
| C5H8#+CH3=C5H7#+CH4 | $2.0\ x10^{11}$ | 0.0 | 9.6 | Estimated [e] | (19) |
| C5H8#+CH3=C5H7#Y+CH4 | $2.0\ x10^{11}$ | 0.0 | 7.3 | Estimated [d] | (20) |
| C5H8#+CH3=C5H7#V+CH4 | 1.96 | 3.5 | 11.7 | Estimated [d] | (21) |
| C5H8#+O=C5H7#+R2OH | $2.6x10^{13}$ | 0.0 | 5.2 | Estimated [e] | (22) |
| C5H8#+O=C5H7#Y+R2OH | $1.7x10^{11}$ | 0.7 | 3.2 | Estimated [d] | (23) |
| C5H8#+O=C5H7#V+R2OH | $1.2x10^{11}$ | 0.7 | 7.6 | Estimated [d] | (24) |
| C5H8#+C2H5=C5H7#+C2H6 | $2.0x10^{11}$ | 0.0 | 11.0 | Estimated [e] | (25) |
| C5H8#+C2H5=C5H7#Y+C2H6 | 8.8 | 3.5 | 4.1 | Estimated [d] | (26) |
| **Reactions of cyclopentenyl and derived radicals** | | | | | |
| C5H7#=C5H7#Y | $2.3x10^{11}$ | 1.0 | 44.1 | Estimated[f] | (27) |
| C5H7#V=C5H7#Y | $6.7x10^{12}$ | 0.0 | 49.5 | Estimated[g] | (28) |
| C5H7#=C5H7-1s | $2.0\ x10^{13}$ | 0.0 | 35.5 | Estimated [d] | (29) |
| C5H7-1s=C2H2+C3H5Y | $2.0\ x10^{13}$ | 0.0 | 33.0 | Estimated [d] | (30) |
| C5H7#Y=C5H7Y | $1.3x10^{13}$ | 0.0 | 35.0 | Estimated [d] | (31) |
| C5H7-1s=C5H7Y | $3.4x10^{13}$ | 0.0 | 34.5 | Estimated [d] | (32) |
| C5H7-2s=C5H7Y | $1.3x10^{13}$ | 0.0 | 44.5 | Estimated [d] | (33) |
| C5H7-3t=C5H7Y | $5.0x10^{13}$ | 0.0 | 37.0 | Estimated [d] | (34) |
| C5H7-4s=C5H7Y | $2.0x10^{13}$ | 0.0 | 47.0 | Estimated [d] | (35) |
| C5H7-2s=C2H3V+aC3H4 | $2.0x10^{13}$ | 0.0 | 35.5 | Estimated [d] | (36) |
| C5H7-3t=CH3+C4H4 | $2.0x10^{13}$ | 0.0 | 31.5 | Estimated [d] | (37) |
| C5H7-4s=C2H3V+pC3H4 | $2.0x10^{13}$ | 0.0 | 31.0 | Estimated [d] | (38) |
| C5H7#V=C5H7-5p | $2.0x10^{13}$ | 0.0 | 33.0 | Estimated [d] | (39) |
| C5H7#V=C5H7-12-5p | $2.0x10^{13}$ | 0.0 | 31.0 | Estimated [d] | (40) |
| C5H7-5p=C2H4+C3H3 | $2.0x10^{13}$ | 0.0 | 22.5 | Estimated [d] | (41) |
| C5H7-12-5p=C2H4+C3H3 | $2.0x10^{13}$ | 0.0 | 35.5 | Estimated [d] | (42) |
| C5H7#=C5H6#+H | $6.4x10^{13}$ | 0.0 | 34.8 | Estimated [d] | (43) |
| C5H7#Y=C5H6#+H | $3.0x10^{13}$ | 0.0 | 50.5 | Estimated [d] | (44) |
| C5H7#+O2=C5H6#+HO2 | 5.2E11 | 0.0 | 2.5 | Estimated[h] | (45) |
| C5H7#Y+O2=C5H6#+HO2 | 1.6x10$^{12}$ | 0.0 | 15.2 | Estimated[h] | (46) |
| C5H7#Y+HO2=C2H4+CH2CHCO+OH | 1.0x10$^{13}$ | 0.0 | 0.0 | Estimated[i] | (47) |
| C5H7#Y+R1H=C5H6#+H2 | 1.8x10$^{13}$ | 0.0 | 0.0 | Estimated[i] | (48) |
| C5H7#Y+R2OH=C5H6#+H2O | 6.0x10$^{12}$ | 0.0 | 0.0 | Estimated[i] | (49) |
| C5H7#Y+CH3=MCP | 5.0x10$^{12}$ | 0.0 | 0.0 | Estimated[j] | (50) |
| **Reactions of cyclopentyl and derived radicals** | | | | | |
| C5H9#=C5H9 | 2.0x10$^{12}$ | 0.005 | 36.5 | SIRJEAN06[30] | (51) |
| C5H9=C5H9Y | 3.3x10$^{9}$ | 1.0 | 32.3 | estimated[f] | (52) |
| C5H9=C3H5Y+C2H4 | 3.3x10$^{13}$ | 0.0 | 22.5 | Estimated[d] | (53) |
| C5H9=H+C5H8 | 3.0x10$^{13}$ | 0.0 | 38.0 | Estimated [d] | (54) |
| C5H9+O2=>C5H8+HO2 | 1.6x10$^{12}$ | 0.0 | 5.0 | estimated[e] | (55) |
| **Reactions of methylcyclopentene and derived radicals** | | | | | |
| MCP+H=RMCP1+H2 | 9.0x10$^{6}$ | 2.0 | 5.0 | Estimated [e] | (56) |
| MCP+H=RMCPY1+H2 | 5.0x10$^{4}$ | 2.5 | -1.9 | Estimated [d] | (57) |
| MCP+H=RMCP2+H2 | 2.9x10$^{7}$ | 2.0 | 7.7 | Estimated [e] | (58) |
| MCP+H=RMCPY2+H2 | 2.5x10$^{4}$ | 2.5 | -2.8 | Estimated [g] | (59) |
| MCP+OH=RMCP1+H2O | 2.6x10$^{6}$ | 2.0 | -0.8 | Estimated [e] | (60) |
| MCP+OH=RMCPY1+H2O | 3.0x10$^{6}$ | 2.0 | -1.5 | Estimated [d] | (61) |
| MCP+OH=RMCP2+H2O | 2.7x10$^{6}$ | 2.0 | 0.4 | Estimated [e] | (62) |
| MCP+OH=RMCPY2+H2O | 1.3x10$^{6}$ | 2.0 | -2.6 | Estimated [d] | (63) |
| RMCP1=C2H2+C4H7Y | 2.0x10$^{13}$ | 0.0 | 33.0 | Estimated [d] | (64) |
| RMCP1=CH3+C5H6# | 2.0x10$^{13}$ | 0.0 | 31.0 | Estimated [e] | (65) |
| RMCP2=C2H4+nC4H5 | 2.0x10$^{13}$ | 0.0 | 28.7 | Estimated [e] | (66) |
| RMCP1+O2=MCPD+HO2 | 2.6x10$^{11}$ | 0.0 | 2.5 | Estimated[h] | (69) |
| RMCPY1+O2=MCPD+HO2 | 1.6x10$^{12}$ | 0.0 | 15.2 | Estimated[h] | (70) |
| RMCPY1+H=MCP | 1.0x10$^{14}$ | 0.0 | 0.0 | Estimated [a] | (71) |
| RMCPY2+H=MCP | 1.0x10$^{14}$ | 0.0 | 0.0 | Estimated [a] | (72) |
| **Reactions of methylcyclopentadiene and derived radicals** | | | | | |
| C5H5#+CH3=MCPD | 0.5x10$^{13}$ | 0.0 | 0.0 | Estimated [j] | (73) |
| MCPD+H=RMCP1 | 2.6x10$^{13}$ | 0.0 | 3.2 | Estimated [d] | (74) |
| MCPD+H=RMCPY1 | 2.6x10$^{13}$ | 0.0 | 3.2 | Estimated [d] | (75) |
| MCPD+H=RMCPD+H2 | 2.9x10$^{7}$ | 2.0 | 7.7 | Estimated [e] | (76) |
| MCPD+OH=RMCPD+H2O | 2.7x10$^{6}$ | 2.0 | 0.4 | Estimated [e] | (77) |
| MCPD+H=RMCPDY+H2 | 2.5 x10$^{4}$ | 2.5 | -2.7 | Estimated [g] | (78) |
| MCPD+OH=RMCPDY+H2O | 1.3x10$^{6}$ | 2.0 | -2.7 | Estimated [g] | (79) |
| RMCPDY=RMCPD | 3.0x10$^{12}$ | 0.0 | 50.4 | LIFSHITZ05[8] | (80) |
| RMCPD=C6H7# | 1.4x10$^{13}$ | 0.0 | 17.4 | LIFSHITZ05 | (81) |
| RMCPDY+H=MCPD | 1.0x10$^{14}$ | 0.0 | 0.0 | Estimated [a] | (82) |
| **Reaction leading to toluene** | | | | | |
| C5H5#+C2H2=>benzyl | 1.0x10$^{12}$ | 0.0 | 7.0 | Estimated [k] | (83) |

[a] : Rate constant taken equal to that of the recombination of H atoms with alkyl radicals as proposed by Allara et al. [31].

[b]: The rate constant of this bimolecular initiation with oxygen molecule has been calculated as proposed by Ingham et al [32]: A is taken equal to n × 7x10$^{12}$ $cm^3.mol^{-1}s^{-1}$, where n is the number of alkyllic hydrogen atoms abstractable, to n × 7x10$^{11}$ $cm^3$ $mol^{-1}$ $s^{-1}$, where n is the number of allyllic hydrogen atoms abstractable or to n × 1x10$^{12}$ $cm^3.mol^{-1}s^{-1}$, where n is the number of vinylic hydrogen atoms abstractable, and the activation energy to the reaction enthalpy.

[c]: Rate constant multiplied by 4 compared to the value estimated by using the correlations proposed by Heyberger et al. [33] in the case of alkenes.

[d]: Rate constant estimated by using the correlations proposed by Heyberger et al. [33, 35] in the case of alkenes.

[e]: Rate constant estimated by using the correlations proposed by Buda et al. [34] in the case of alkanes.

[f]: The rate constant for this isomerisation has been calculated by using the method described in the part II of this paper [10], with an increase of 5 kcal/of the activation energy to take into account the formation of a bicyclic transition state in the case of reaction 27.

[g]: Rate constant estimated by using the correlations proposed by Heyberger et al. [34, 35] in the case of alkenes, with an increase of 5 kcal/of the activation energy to take into account the formation of a bicyclic transition state.

[h]: Rate constant estimated by using the correlations proposed by Touchard et al. [36] in the case of 1-pentene and 1-hexene.

[i]: Rate constant estimated by analogy with the values proposed by Tsang [37] for allyl radicals.

[j] : Rate constant of this recombination calculated by the modified collision theory at 1500 K using software KINGAS [38].

[k]: see text.

**TABLE II: NAMES, FORMULAE AND HEATS OF FORMATION FOR $C_5$ AND $C_6$ SPECIES INVOLVED IN TABLE II.**

The heats of formation have been calculated by software THERGAS [19] at 300 K in kcal.mol$^{-1}$.

| Species | Structure | $\Delta H_f$ (300K) | Species | Structure | $\Delta H_f$ (300K) |
|---|---|---|---|---|---|
| Cyclopentene (C5H8#) | 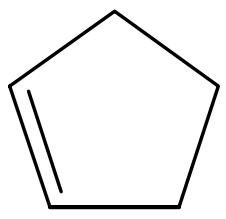 | 8.31 | Cyclopentadiene (C5H6#) | 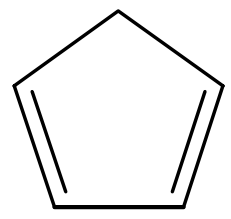 | 32.0 |
| Cyclopentyl radical (C5H9#) | 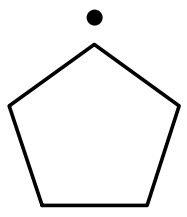 | 23.9 | Alkylic cyclopentenyl radical (C5H7#) | 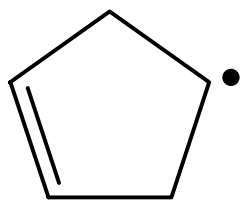 | 50.7 |
| Allylic cyclopentenyl radical (C5H7#Y) | 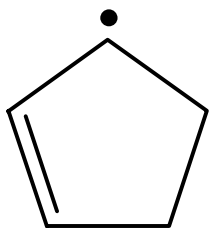 | 38.5 | Vinylic cyclopentenyl radical (C5H7#V) | 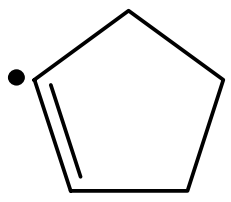 | 64.9 |
| 1,4-pentadien-1-yl radical (C5H7-1s) | 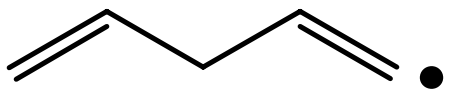 | 84.2 | 1,3-pentadien-5-yl or 1,4-pentadien-3-yl radical (C5H7Y) | 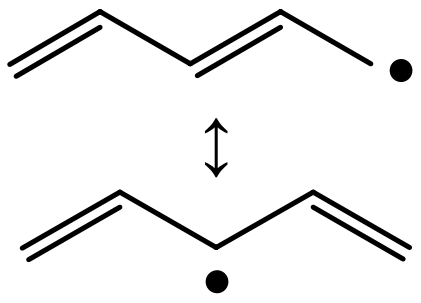 | 48.9 |
| 1,4-pentadien-2-yl radical (C5H7-2s) | 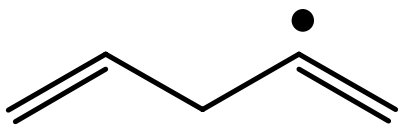 | 81.8 | 1,3-pentadien-3-yl radical (C5H7-3t) | 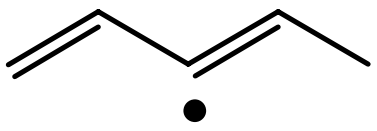 | 74.6 |
| 1,3-pentadien-4-yl (C5H7-4s) | 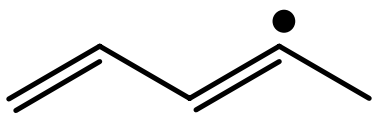 | 74.9 | 1-pentyn-5yl (C5H7-5p) | 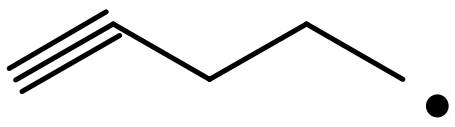 | 83.7 |
| 1,2-pentadien-5-yl (C5H7-12-5p) | 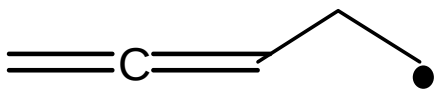 | 82.9 | 1-penten-5yl (C5H9) | 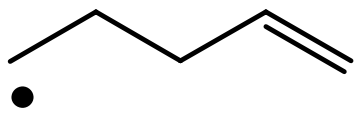 | 43.7 |
| 1-penten-3yl (C5H9Y) | 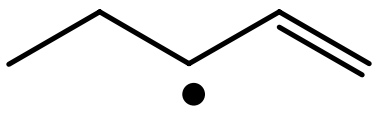 | 26.4 | 1,2-pentadiene (C5H8-12) | 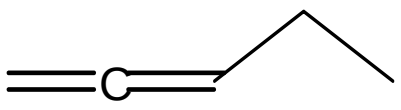 | 34.1 |
| 1,3-pentadiene (C5H8) | 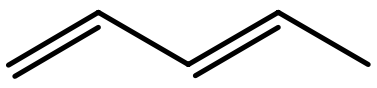 | 18.7 | Methylcyclopentene (MCP) | 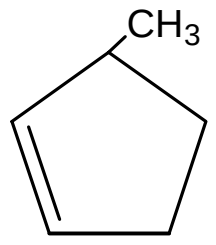 | 2.27 |
| Methylcyclopentadiene (MCPD) | 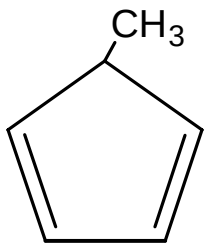 | 25.0 | 3-methyl cyclopenten-4yl (RMCP1) | 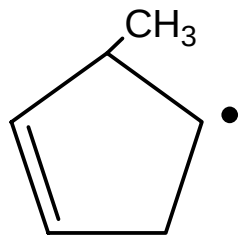 | 44.7 |
| 3-methyl cyclopenten-5yl (RMCPY1) | 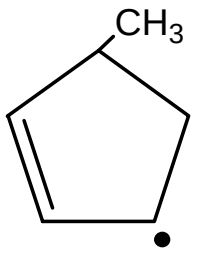 | 32.5 | Cyclopentenylmethylene (RMCP2) | 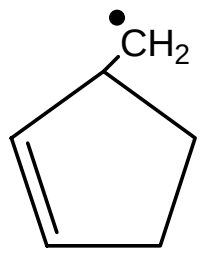 | 50.4 |
| 3-methyl cyclopenten-3yl (RMCPY2) | 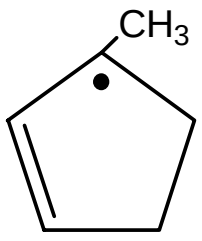 | 32.5 | Methyl cyclopentadienyl (RMCPDY) | 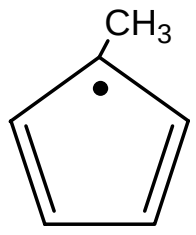 | 56.8 |
| Cyclopentadienyl methylene (RMCPD) | 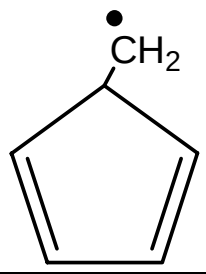 | 73.1 | | | |

**FIGURE CAPTIONS**

Figure 1: Typical chromatogram of $C_1$-$C_7$ compounds obtained at a distance of 0.59 cm from the burner (oven temperature program: 313 K during 22 min, then a rise of 1 K/min until 523 K).

Figure 2: Temperature profiles in the three flames: experimental measurements performed without and with the sampling probe and profile used for simulation.

Figure 3: Profiles of the mole fractions of oxygen and $C_1$ species. Points are experiments and lines simulations. Full lines correspond to the flame seeded with cyclopentene and broken lines to a simulated flame of pure methane at $\Phi$=1.79 (see text). The broken line can nearly not be seen in fig. 3a.

Figure 4: Profiles of the mole fractions of hydrogen and $C_2$ species. Points are experiments and lines simulations. Full lines correspond to the flame seeded with cyclopentene and the broken lines to a simulated flame of pure methane at $\Phi$=1.79 (see text).

Figure 5: Profiles of the mole fractions $C_3$ species. Points are experiments and lines simulations. Full lines correspond to the flame seeded with cyclopentene and the broken lines to a simulated flame of pure methane at $\Phi$=1.79 (see text).

Figure 6: Profiles of the mole fractions $C_4$ species. Points are experiments and lines simulations.

Figure 7: Profiles of the mole fractions $C_5$ species and aromatic compounds. Points are experiments and lines simulations. Full lines correspond to simulations with the full mechanism, dotted lines to the mechanism without reaction 83 and thin broken lines to the mechanism without reaction 81.

Figure 8: Flow rate analysis for the consumption of the cyclopentene for a distance of 0.36 cm

from the burner corresponding to a temperature of 990 K and a conversion of 47 % of cyclopentene. The size of the arrows is proportional to the relative flow rates.

Figure 9: Simulated mole fractions of (a) H-atoms and methyl radicals (full lines correspond to the flame seeded with cyclopentene and broken lines to a simulated flame of pure methane at $\Phi$=1.79) and of (b) cyclohexadiene and other important cyclic compounds.

Figure 10: Flow rate analysis for the formation and consumption of benzene and toluene for a distance of 0.48 cm from the burner corresponding to a temperature of 1350 K, a conversion of 90% of cyclopentene and close to the peak of benzene profile. The size of the arrows is proportional to the relative flow rates.

Figure 1

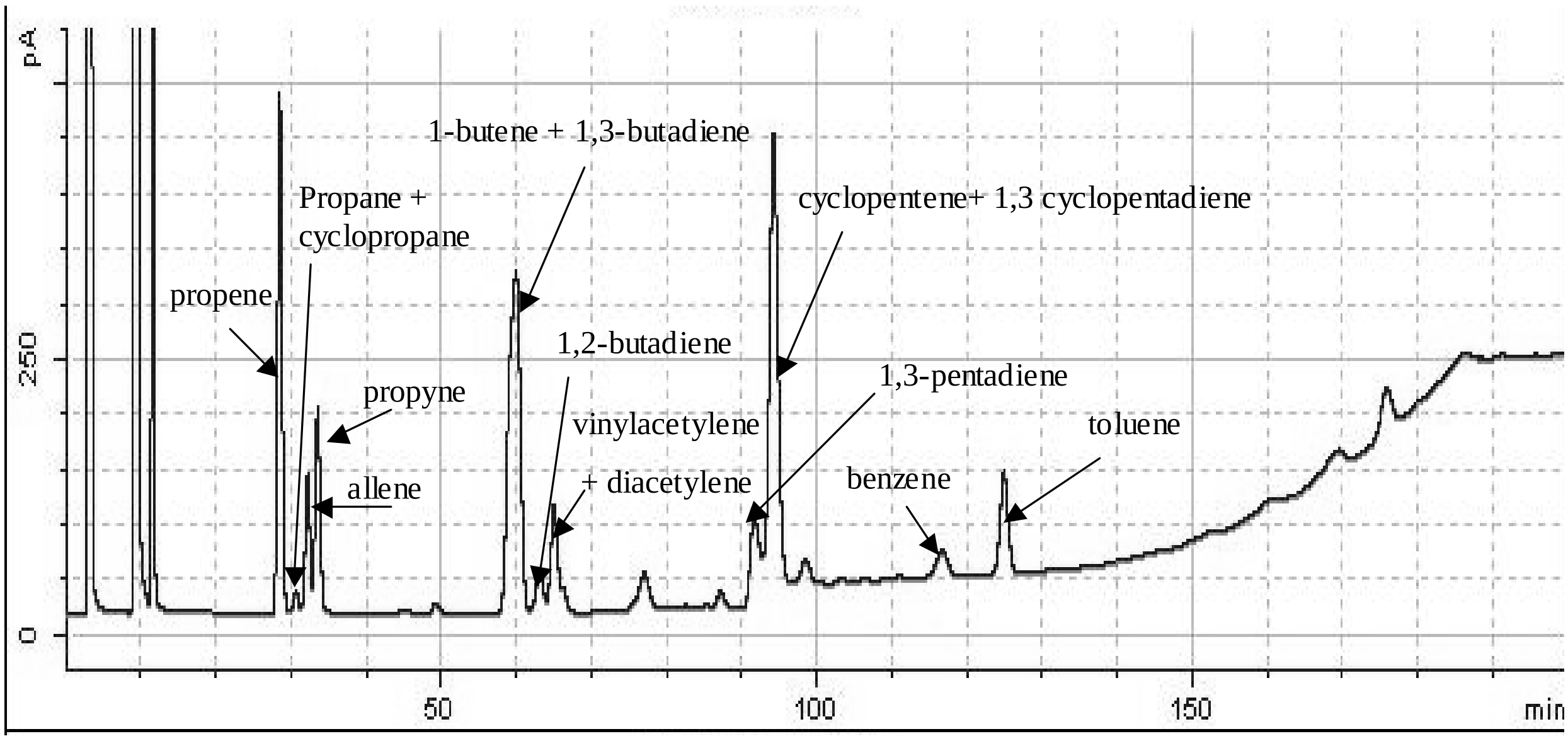

Figure 2

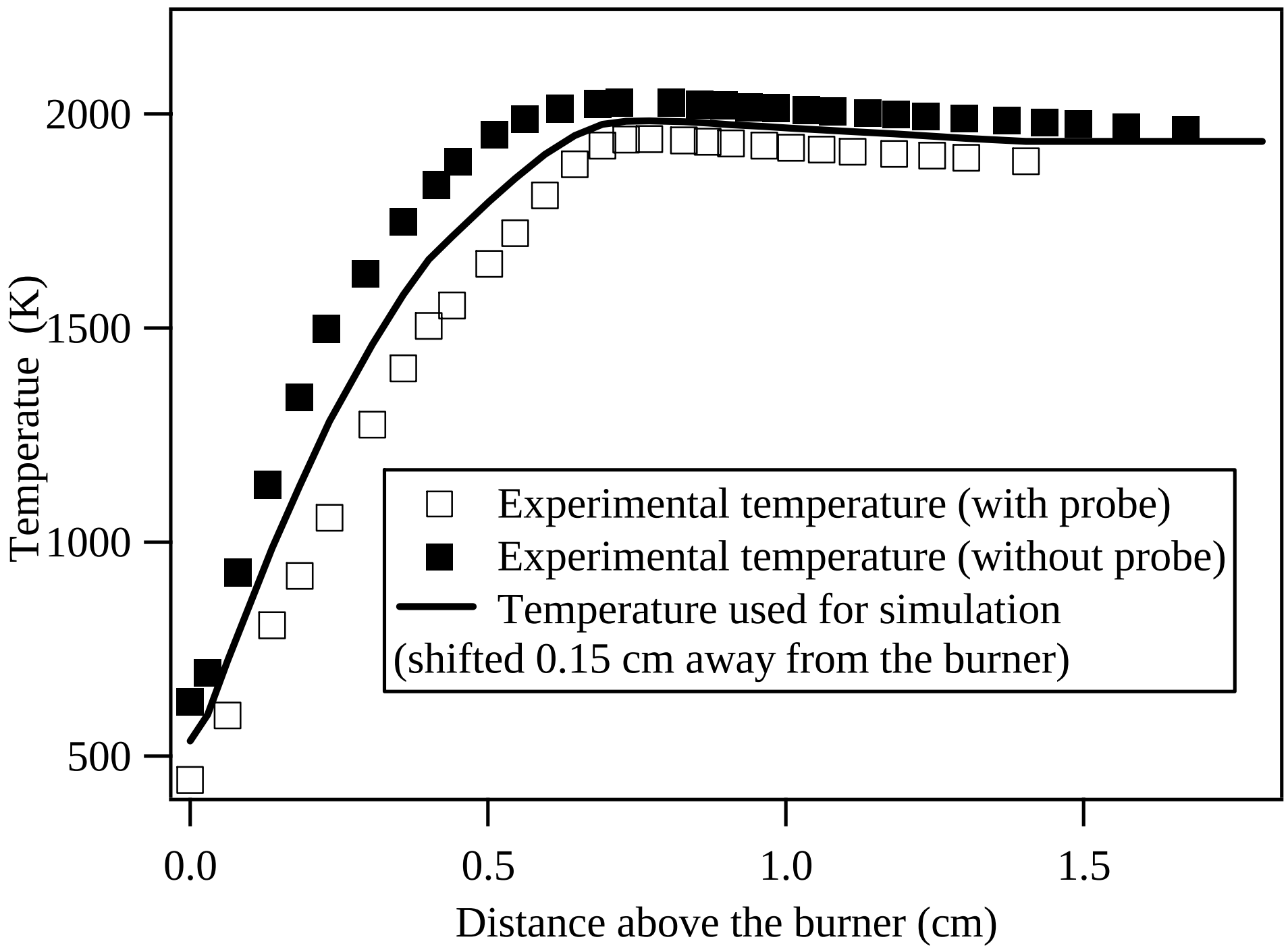

2000
1500
1000
500
Temperatue (K)
0.0
0.5
1.0
1.5
Distance above the burner (cm)
Experimental temperature (with probe)
Experimental temperature (without probe)
Temperature used for simulation
(shifted 0.15 cm away from the burner)

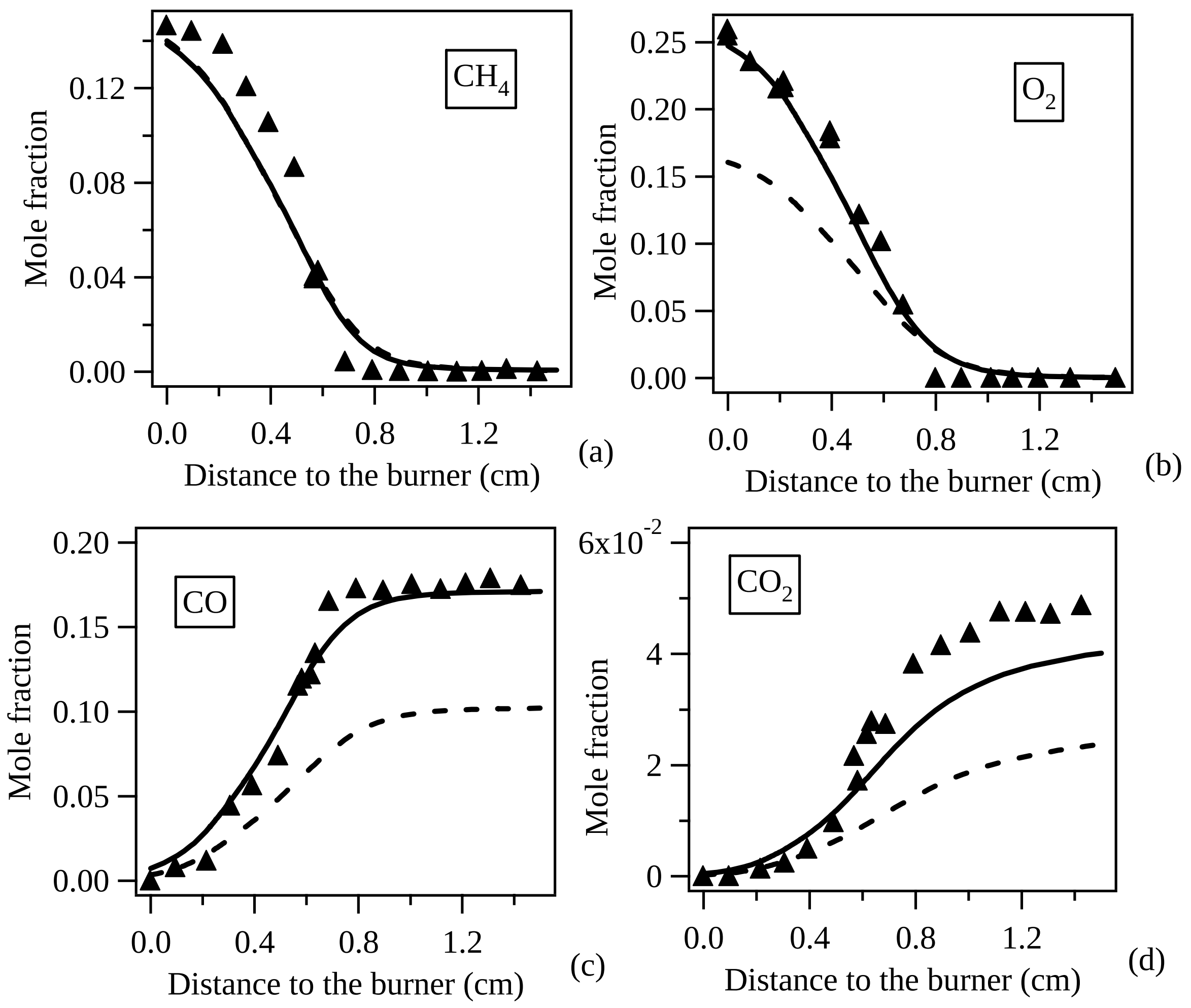

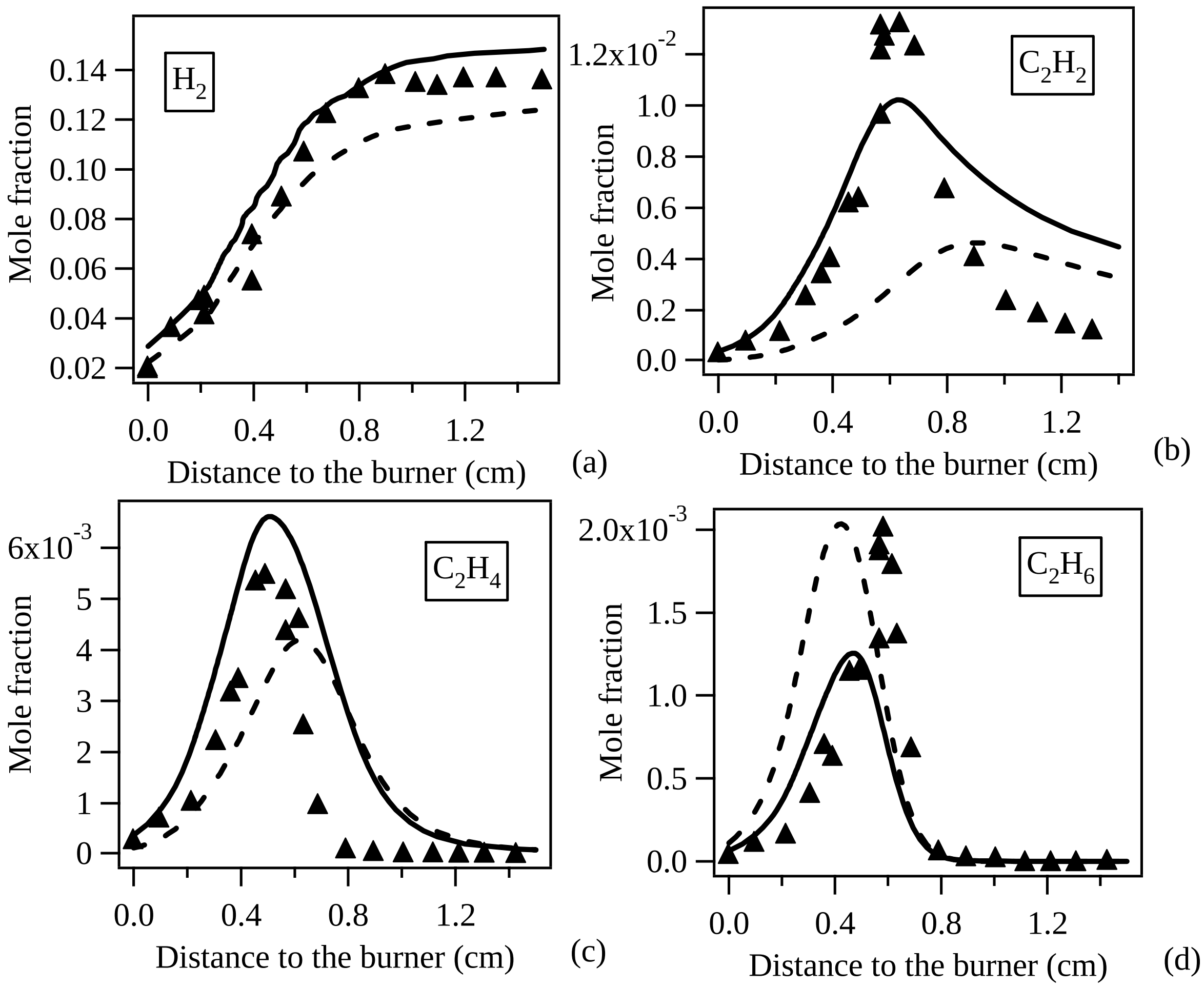

Figure 5

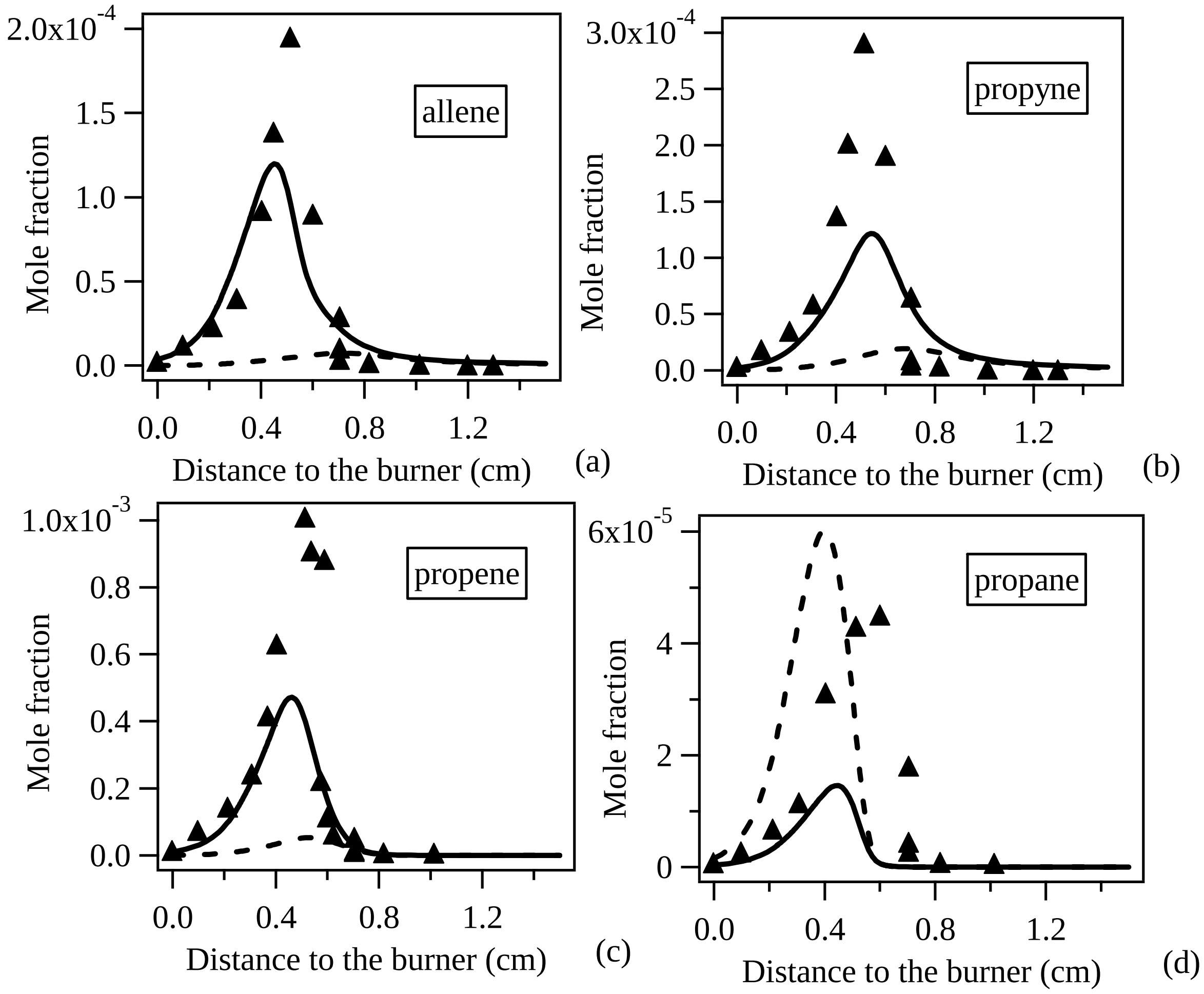

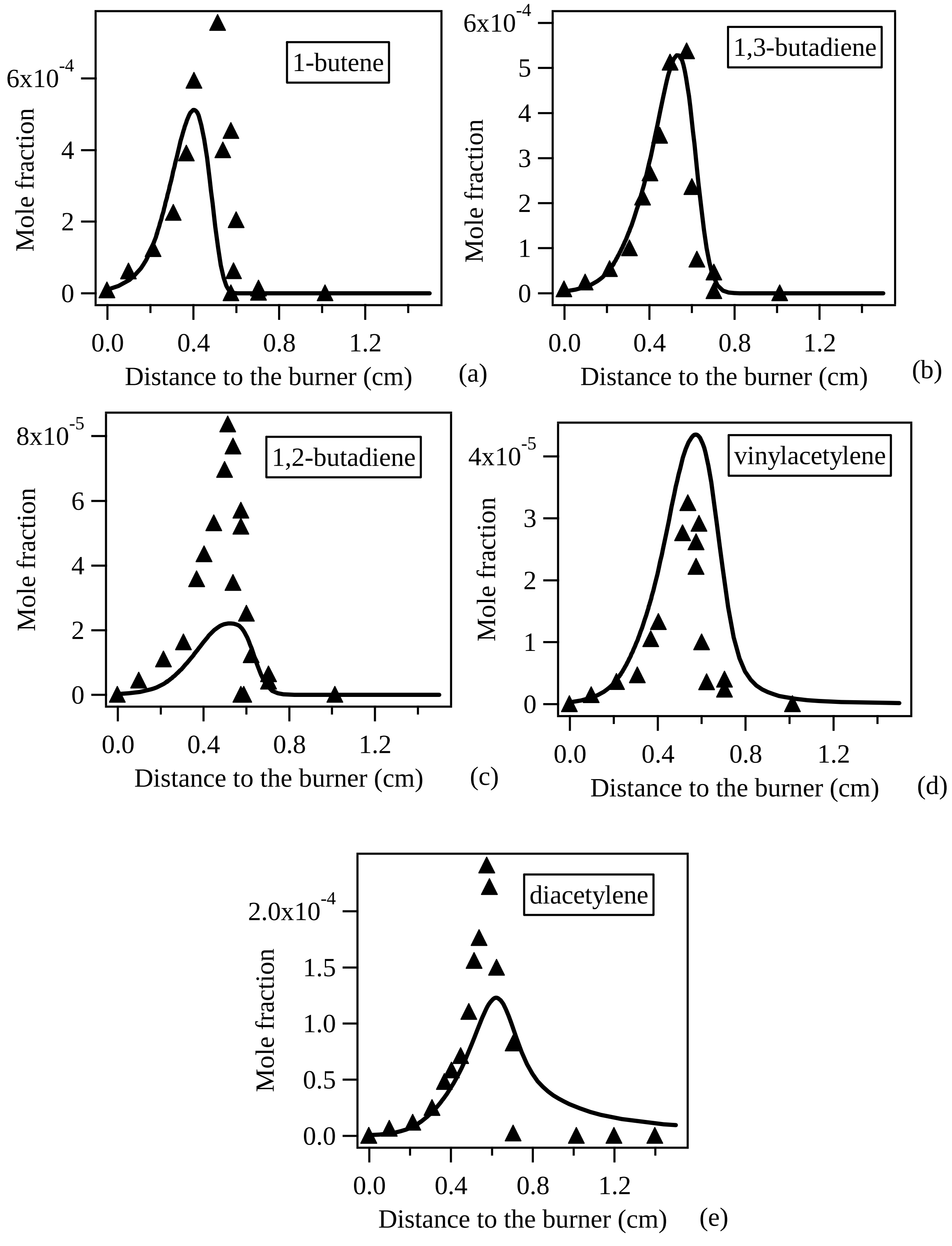

Figure 7

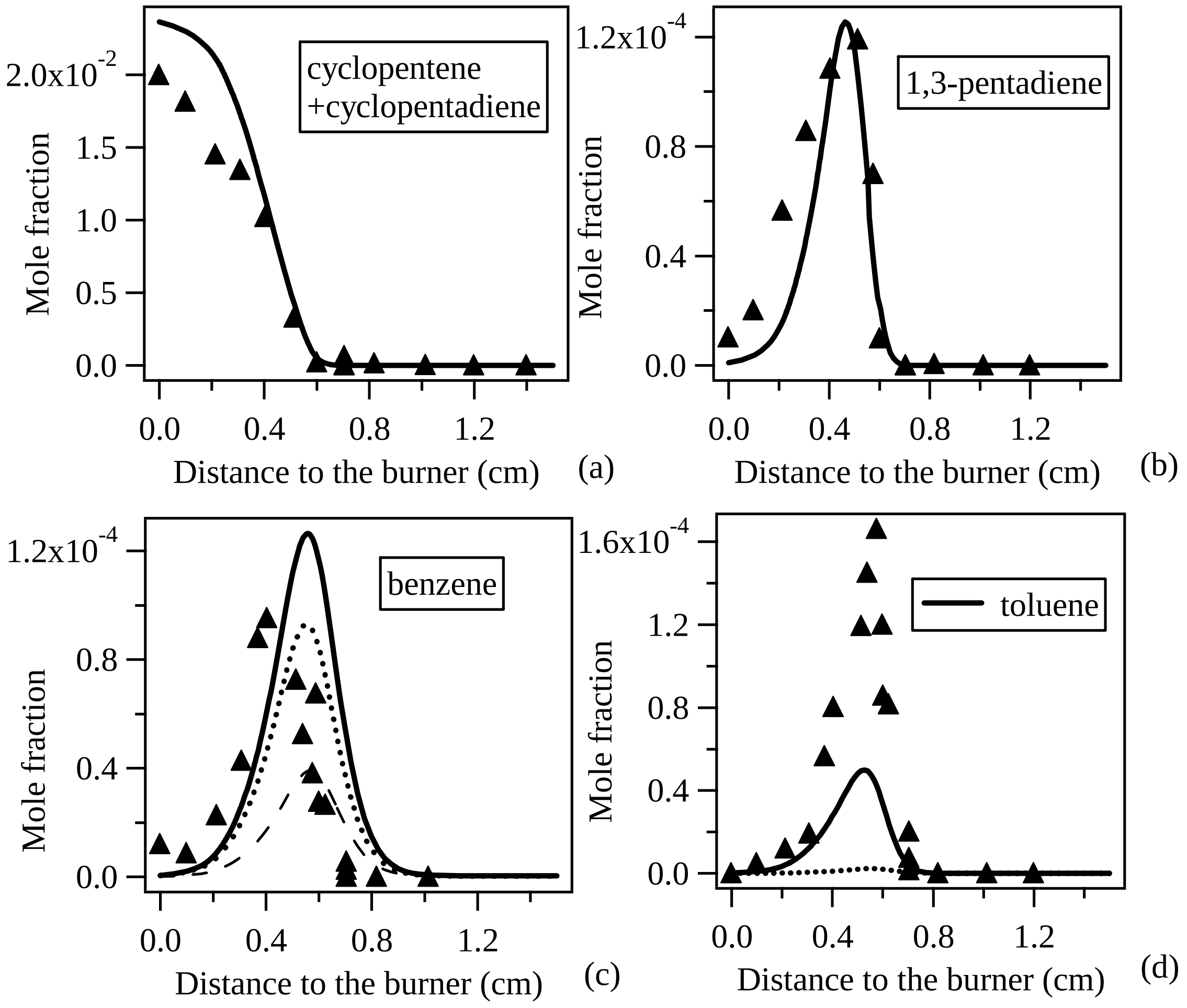

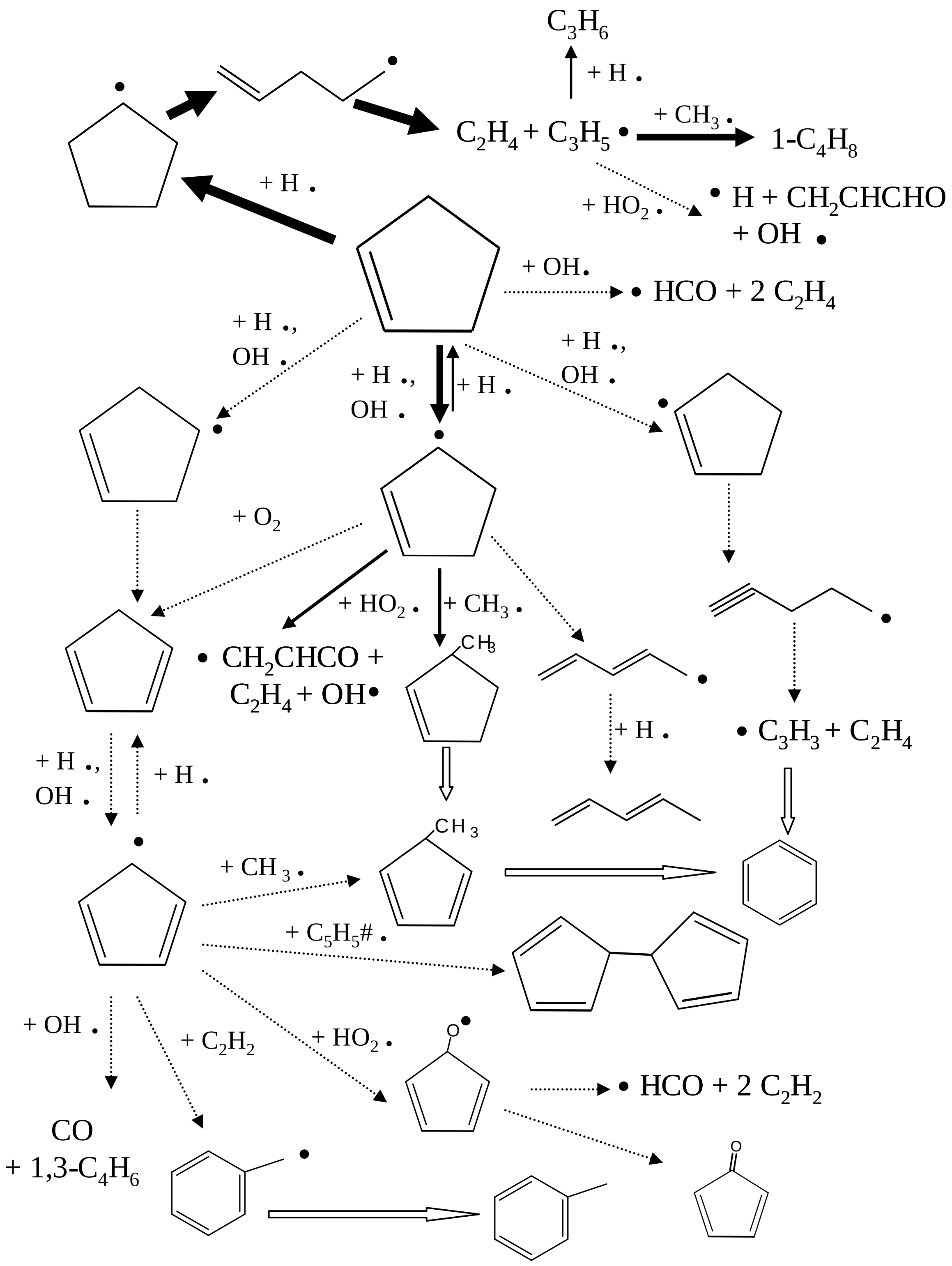

Figure 9

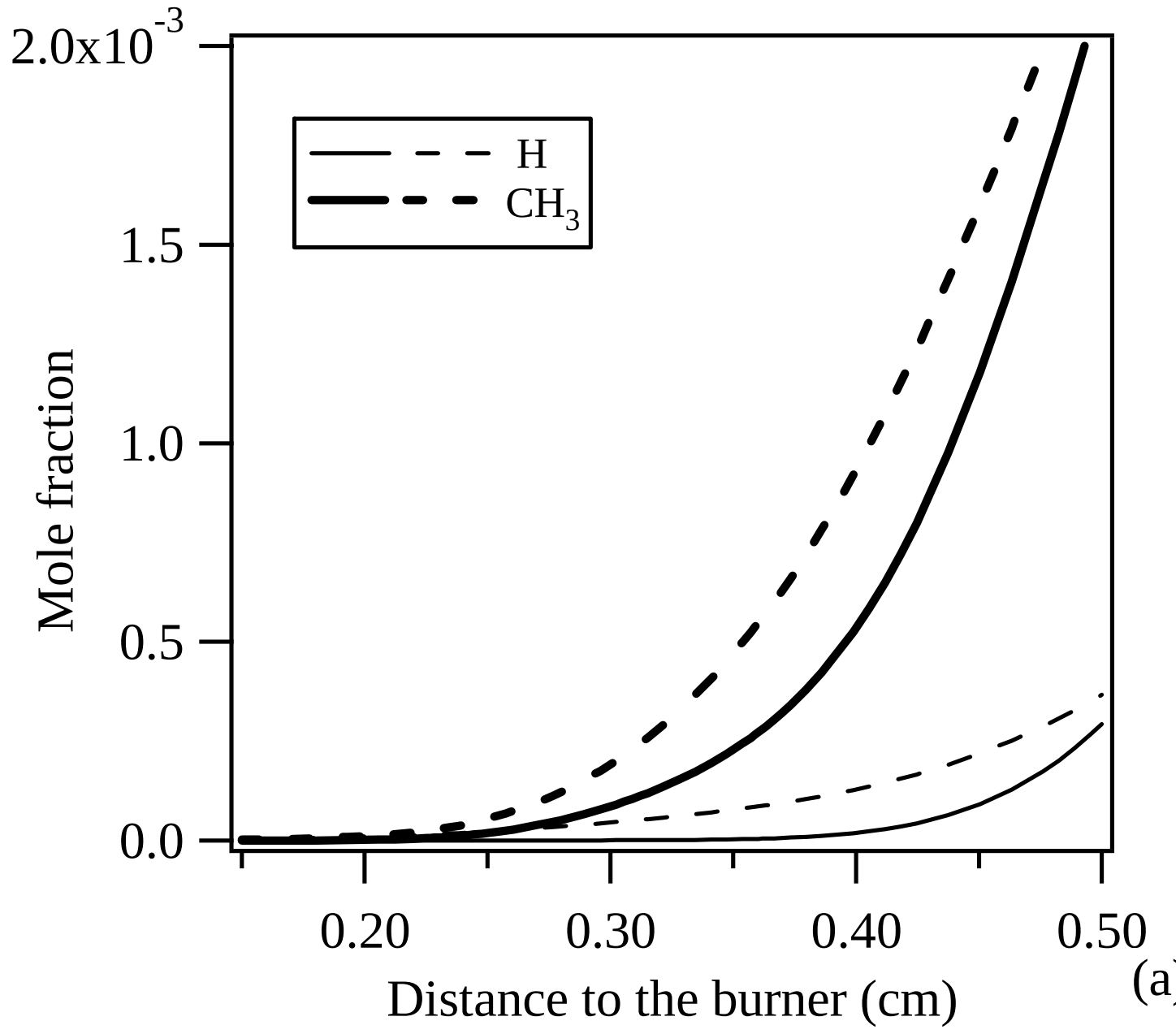

(a)

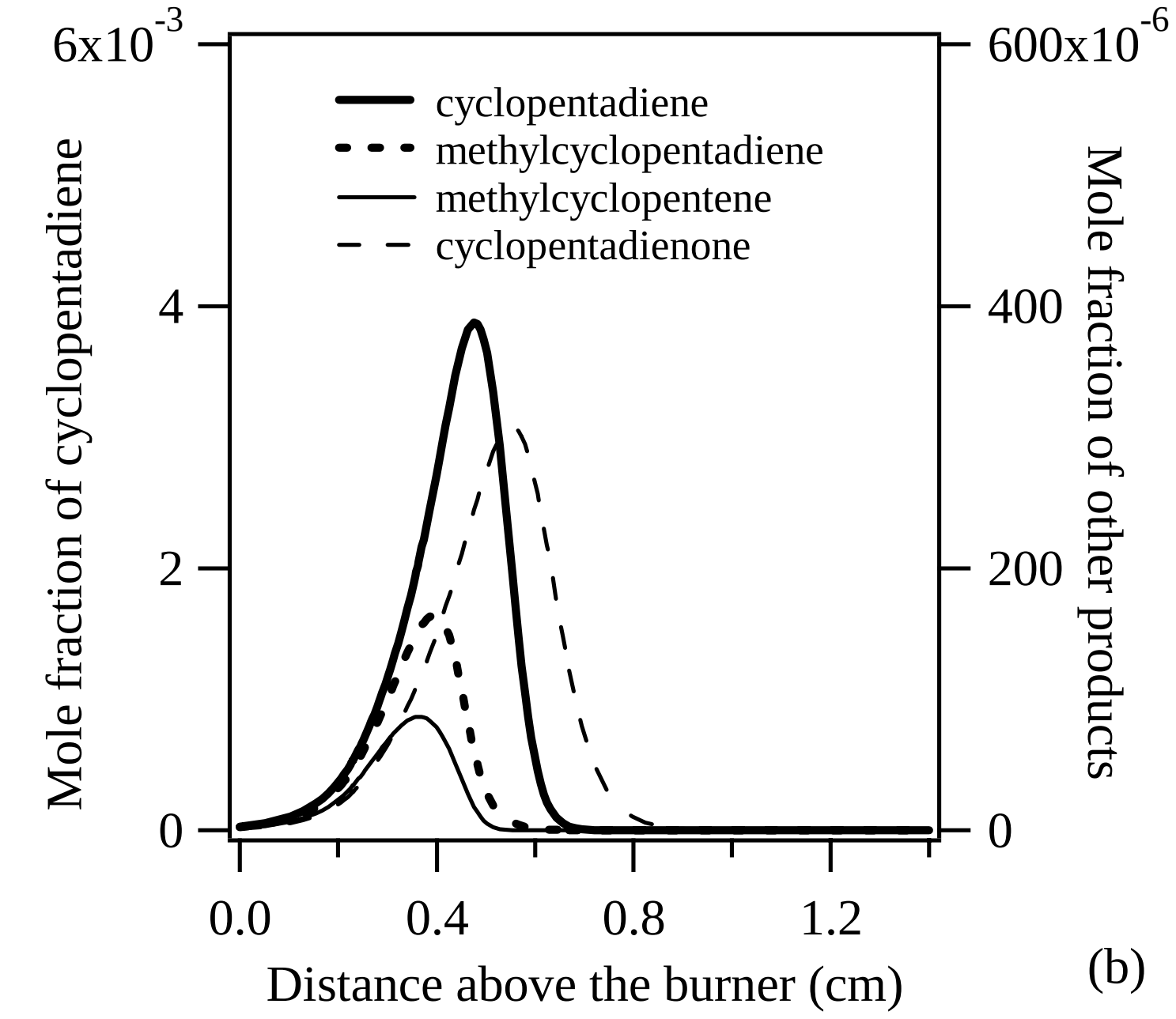

(b)

Figure 10

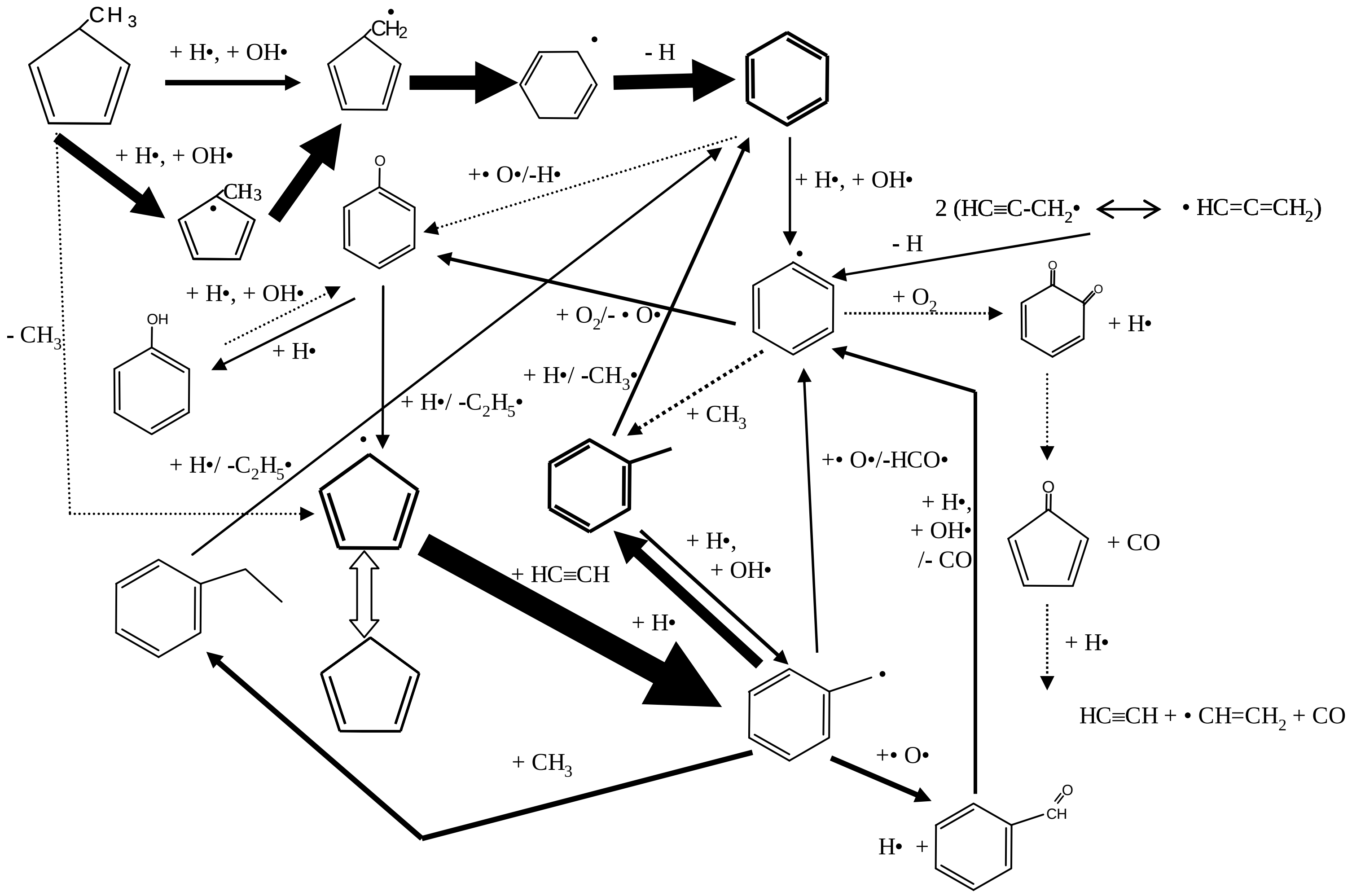

+ H•, + OH•
- H
+ H•, + OH•
+• O•/-H•
+ H•, + OH•
2 (HC≡C-CH2• ⟷ • HC=C=CH2)
- H
+ H•, + OH•
+ H•
- CH3
+ O2/- • O•
+ O2
+ H•
+ H•/ -CH3•
+ H•/ -C2H5•
+ CH3
+ H•/ -C2H5•
+• O•/-HCO•
+ H•, + OH• /- CO
+ CO
+ H•,
+ OH•
+ HC≡CH
+ H•
+ H•
HC≡CH + • CH=CH2 + CO
+• O•
+ CH3
H• +